# Unsteady Shallow Meandering Flows in Rectangular Reservoirs: A Modal Analysis of URANS Modelling

Daniel Valero[a], Daniel B. Bung[b], Sebastien Erpicum[c], Yann Peltier[d] & Benjamin Dewals[e]

[a]*Institute for Water and River Basin Development (IWG), Karlsruhe Institute of Technology (KIT), Engesserstraße 22, 76131 Karlsruhe, Germany*

*Water Resources and Ecosystems, IHE Delft Institute for Water Education, Westvest 7, 2611 AX Delft, the Netherlands*

*Formerly: Hydraulic Engineering Section (HES), Aachen University of Applied Sciences (FH Aachen), Bayernallee 9, 52066 Aachen, Germany*

*corresponding author: daniel.valero@kit.edu*

[b]*Hydraulic Engineering Section (HES), Aachen University of Applied Sciences (FH Aachen), Bayernallee 9, 52066 Aachen, Germany*

*e-mail: bung@fh-aachen.de*

[c]*Hydraulics in Environmental and Civil Engineering (HECE), University of Liège (ULiège), Allée de la découverte 9 - Bâtiment B52, 4000 Liège, Belgium*

*e-mail: s.erpicum@uliege.be*

[d]*Artelia, Agence de Lille, 300 Rue de Lille, 59520 Marquette-lez-Lille, France*

*Formerly: Hydraulics in Environmental and Civil Engineering (HECE), University of Liège (ULiège), Allée de la découverte 9 - Bâtiment B52, 4000 Liège, Belgium*

*e-mail: yann.peltier90@gmail.com*

[e]*Hydraulics in Environmental and Civil Engineering (HECE), University of Liège (ULiège), Allée de la découverte 9 - Bâtiment B52, 4000 Liège, Belgium*

*e-mail: b.dewals@ulg.ac.be*

## Abstract

Shallow flows are common in natural and human-made environments. Even for simple rectangular shallow reservoirs, recent laboratory experiments show that the developing flow fields are particularly complex, involving large-scale turbulent structures. For specific combinations of reservoir size and hydraulic conditions, a meandering jet can be observed. While some aspects of this pseudo-2D flow pattern can be reproduced using a 2D numerical model, new 3D simulations, based on the unsteady Reynolds-Averaged Navier-Stokes equations, show consistent advantages as presented herein. A Proper Orthogonal Decomposition was used to characterize the four most energetic modes of the meandering jet at the free surface level, allowing comparison against experimental data and 2D (depth-averaged) numerical results. Three different isotropic eddy viscosity models (RNG $k$-$\varepsilon$, $k$-$\varepsilon$, $k$-$\omega$) were tested. The 3D models accurately predicted the frequency of the modes, whereas the amplitudes of the modes and associated energy were damped for the friction-dominant cases and augmented for non-frictional ones. The performance of the three turbulence models remained essentially similar, with slightly better predictions by RNG $k$-$\varepsilon$ model in the case with the highest Reynolds number. Finally, the Q-criterion was used to identify vortices and study their dynamics, assisting on the identification of the differences between: i) the three-dimensional phenomenon (here reproduced), ii) its two-dimensional footprint in the free surface (experimental observations) and iii) the depth-averaged case (represented by 2D models).

**Keywords**: *coherent structures; hydraulic modelling; model performance; Proper Orthogonal Decomposition; Q-criterion; Unsteady Reynolds Averaged Navier Stokes*

## 1. Introduction

Rectangular shallow reservoirs are hydraulic structures commonly used as storm basins, sedimentation tanks, service reservoirs, aquaculture ponds, among other applications (Liu et al., 2013; Oca and Masaló, 2007; Zhang et al., 2014). Complex turbulent flow fields occur in such reservoirs, involving symmetric, asymmetric or meandering jets (Stovin and Saul, 2000; Dewals et al., 2008; Kantoush et al., 2008; Dufresne et al., 2010a; Camnasio et al., 2011; Peltier et al., 2014a). These flow characteristics directly affect the operation of the reservoirs. For instance, the amount and location of sediment deposits strongly depends on the flow pattern (Adamsson et al., 2003; Dufresne et al., 2010b; Sébastian et al., 2015; Isenmann et al., 2017).

The hydraulic and geometric conditions leading to symmetric and asymmetric jets were extensively studied experimentally (Aloui and Souhar, 2000; Camnasio et al., 2011; Canbazoglu and Bozkir, 2004; Dufresne et al., 2010a; Kantoush et al., 2008; Mullin et al., 2003; Oca and Masaló, 2007), numerically (Camnasio et al., 2013; Dewals et al., 2008; Dufresne et al., 2011; Esmaeili et al., 2016; Khan et al., 2013; Peltier et al., 2015; Peng et al., 2011; Stovin and Saul, 2000) and analytically (Westhoff et al., 2018). In contrast, less attention was given to the case of a meandering jet, despite its engineering relevance. This type of jet is made of large-scale energetic turbulent structures, which enhance the lateral momentum transfer between the jet and the rest of the flow (Chen and Jirka, 1999). As a consequence, meandering jets promote wider lateral spreading of sediment deposits and an overall increase in the reservoir trapping efficiency.

Within the flow, turbulent coherent structures are spatially distributed, articulating the momentum transfer via their characteristic life cycle (Berkooz et al., 1993; Buffin-Bélanger et al., 2013). These macroscale productive features can be understood as the skeleton of turbulence (Franca et al., 2021). The Proper Orthogonal Decomposition (POD) is a tool that can be used to identify these highly energetic structures (Berkooz et al., 1993). So far, only Peltier et al. (2014a; b) performed systematic laboratory experiments to define the conditions of occurrence of meandering jets, and characterize the corresponding flow fields using POD. Similarly, only Peltier et al. (2015) performed computational modelling of meandering jets in rectangular shallow reservoirs and compared the numerical predictions with the experimental observations using the POD analysis. However, the study of Peltier et al. (2015) relies on a 2D shallow-water model (Camnasio et al., 2014; Erpicum et al., 2009), assuming hydrostatic pressure distribution and uniform velocity profiles over the water depth. These simplifications lead to disregarding the possible effects of secondary flow and other 3D flow features.

Other researchers have conducted full 3D simulations of shallow reservoir flows; see for instance the Lake Binaba study of Abbasi et al. (2016), focused on daily-unsteady thermal fluxes where the unsteadiness is driven by daily boundary condition changes. Another example is the 3D numerical analysis of Haun et al. (2013) of the suspended sediment distribution and bed level changes in the Angostura hydropower reservoir in Costa Rica. Seasonal variability in the study of Haun et al. (2013) was represented by a constant water level and inflow discharge and, hence, turbulent-unsteady flows were not directly addressed. Haun and Olsen (2012) conducted a 3D numerical modelling of a flushing operation in the same reservoir, with detailed time resolution, yet for a shorter period than Haun et al. (2013). Haun and Olsen (2012) suggested that a 3D approach presents advantages over a 2D one, since secondary currents can be reproduced together with "*vertical recirculation zones and other flow features where the velocity profile is non-logarithmic*". However, no previous 3D numerical study has dealt with flows in shallow reservoirs at a scale where unsteadiness is driven by turbulent instabilities. Can this unsteadiness be reproduced by 3D (time-averaged) numerical models? And, if they can, which is the expected accuracy?

The answers to these questions are not straightforward and, therefore, here we aim at providing new insights into the ability of 3D unsteady Reynolds-Averaged Navier-Stokes (URANS) equations to better predict the unsteady properties of meandering jets in rectangular shallow reservoirs. As a case study, we used the same four flow conditions as experimentally studied by Peltier et al. (2014b) – which served for the 2D numerical workbench of Peltier et al. (2015). Mesh sensitivity was assessed and three of the nowadays most commonly used turbulence models, namely RNG $k$-$\varepsilon$, $k$-$\varepsilon$ and $k$-$\omega$, were tested.

## 2. Methods

### *2.1. URANS equations*

We used here the URANS equations, which may explicitly reproduce time-dependent flow features as long as the averaging window is considerably smaller than the characteristic time-scale of the unsteadiness (pp. 36-37 of Wilcox, 2006; Rodi, 2017; Spalart, 2000). Among others, Ge and Sotiropoulos (2005) as well as Khosronejad et al. (2012) applied URANS to capture vortical structures in the flow in the vicinity of piers, while Palkin et al. (2016) successfully computed the vortex shedding frequency in the flow detachment downstream of a circular pier, except for low frequency modulations. Here, we embrace a URANS model because it is of utmost engineering interest: most possibly, during the next 10 years, future hydrodynamic studies focused on reservoirs will use URANS modelling instead of more detailed approaches such as Large Eddy Simulations (LES) given the space and time dimensions involved.

The simulations were conducted using the commercial Computational Fluid Dynamics (CFD) package FLOW-3D®. A second order monotonicity preserving explicit advection scheme (Van Leer, 1977) was used, together with an explicit scheme for the viscous terms, and the Generalized Minimal Residual method (GMRES, see projection methods in Saad, 2003) with Krylov subspace dimension 15 (based on authors previous experience, see for instance Valero et al., 2018, or Valero and Bung, 2016 using a smaller Krylov subspace), to solve the sparse linear systems resulting from the Finite Volume Method discretization (Versteeg and Malalasekera, 2007). The Volume of Fluid (VOF) method, as described by Hirt and Nichols (1981), was used for the free surface tracking. Solid boundaries are represented within the mesh using the FAVOR porosity-based technique of Hirt and Sicilian (1985). Smooth wall conditions are assumed in all calculations.

### *2.2. Turbulence closure*

Three turbulence models (RNG $k$-$\varepsilon$, $k$-$\varepsilon$ and $k$-$\omega$) were used here. They are among the most widely used in engineering. They are linear eddy viscosity models – i.e. based on a linear relation between turbulent stresses and velocity gradients (Rodi, 2017). Maximum turbulent length-scale was limited to 7 % of the inlet flow depth of each simulation, which is a typically accepted value in practice (Pope, 2000). This parameter stablishes the minimum energy dissipation, which occurs in regions of lower turbulence. Differently from the study on shallow reservoir instabilities of Dewals et al. (2008), no perturbation is intentionally introduced in the inlet. However, it is believed that minor numerical artefacts can lead to small numerical perturbations introduced in the flow variables. Consequently, small perturbations can be present in the flow even if they are not deliberately introduced. It is expected that these perturbations (of numerical nature) remain small and are only amplified by the flow equations; in the same manner as physically-based appearing perturbations grow (see perturbations´ sensitivity study of Valero et al., 2017).

### *2.3. Case study*

A rectangular shallow reservoir was used as a case study (Fig. 1). The reservoir geometry corresponds to the experimental setup of Peltier et al. (2014b), which was also used in the 2D numerical study of Peltier et al. (2015). The reservoir has a length $L$ = 1 m and a width $B$ = 0.98 m. The inlet channel is 2 m long and $b$ = 0.08 m wide. The outlet channel has the same width as the inlet channel and is 0.13 m long. The corresponding shape factor (Dufresne et al. 2010a) is $\mathrm{SF} = L/\Delta B^{0.6}\, b^{0.4}$ = 4.43; with $\Delta B$ the width of the sudden expansion.

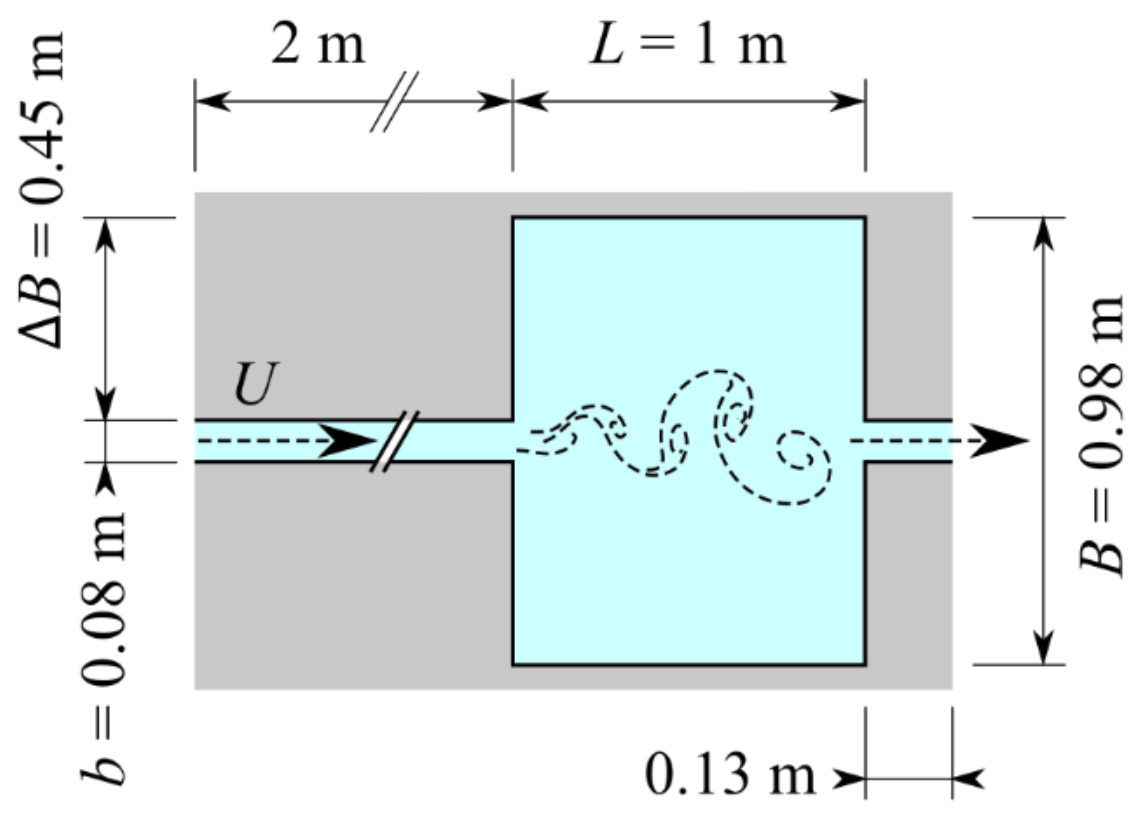

**Figure 1**. Geometry of the considered rectangular shallow reservoir.

The flow rate ($Q$) and flow depth ($H$) were prescribed at the inlet, while only the flow depth was prescribed at the downstream boundary condition. Consistently with Peltier et al. (2014b), four cases of meandering flow were investigated, as detailed in Table 1. They correspond to two different frictional regimes (labelled F and NF, standing respectively for “frictional” and “non frictional”) and two transitional cases in between (labelled FT and NFT).

**Table 1**. Flow parameters of the four considered cases.

| ID | $Q$ ($10^{-3}$ m$^3$/s) | $H$ (cm) | $U$ (m/s) | $\mathsf{F}$ (-) | $\mathsf{R}$ ($10^3$) | $\mathsf{S}$ (-) | Friction regime |
|---|---|---|---|---|---|---|---|
| F | 0.125 | 1.25 | 0.125 | 0.36 | 4.7 | 0.18 | Frictional |
| FT | 0.250 | 1.80 | 0.174 | 0.41 | 8.4 | 0.10 | Frictional close transition |
| NFT | 0.500 | 2.75 | 0.227 | 0.44 | 14.8 | 0.06 | Non-frictional close transition |
| NF | 1.000 | 4.20 | 0.298 | 0.46 | 21.2 | 0.03 | Non-frictional |

In the four cases, the inlet Froude number $\mathsf{F}$ (based on the depth-averaged velocity $U$) corresponds to subcritical flow but remains well above 0.21, which is a critical value for the occurrence of a meandering jet (Peltier et al., 2014a). In contrast, the inlet Reynolds number $\mathsf{R}$ and the frictional number $\mathsf{S}$ (Dufresne et al., 2010a; b) differ substantially between the cases. Cases F and FT have smaller Reynolds numbers compared to cases NFT and NF. Depending on the friction number, contrasting jet behaviours and types of coherent structures are expected due to distinct driving mechanisms, as extensively described by Peltier et al. (2014b), as well as in the supplementary material.

Cell resolution was set to 1 cm per 1 cm in the horizontal plane, whereas the vertical resolution was systematically varied to assess mesh sensitivity (nine levels of refinement, as shown in Table 2). Fixed cell resolution in the horizontal plane aims at simplifying the POD analyses; however, it is still finer (or equivalent) than any previous numerical model conducted in the past for similar flow conditions, and matches the horizontal resolution of Peltier et al. (2015), to which model results are compared.

**Table 2**. Vertical cell resolution for the mesh sensitivity analysis.

| $\Delta z$ (cm) | F | NF |
|---|---|---|
| | RNG $k-\varepsilon$ | RNG $k-\varepsilon$ |
| 0.81 | X | |
| 0.65 | X | |
| 0.54 | X | |
| 0.46 | X | |
| 0.41 | X | X |
| 0.36 | X | |
| 0.33 | X | X |
| 0.30 | X | |
| 0.27 | X | X |

All conducted simulations (26 in total), are composed of a first stage where statistically stationary flow conditions are reached over a simulation time of 180 s. After that time, the flow field is deemed independent of the initial flow conditions. The solution at the last time step of this simulation was then used as the initial condition for a second simulation of 360 s in which velocities were stored with a 25 Hz frequency, which coincides with the experimental sample rate used in the study of Peltier et al. (2014b). This second stage is used to produce data for the subsequent analysis, for which 9,000 velocity fields were extracted near the free surface level (at the first cell immediately underneath). After extraction, in-house Python routines were used to convert velocity fields into the appropriate format to run the POD code of Peltier et al. (2014b), which was applied only inside the shallow reservoir area (without considering the inlet/outlet channel flow).

### *2.4. Proper Orthogonal Decomposition*

Turbulent structures and their respective energy are discriminated through a modal decomposition method (i.e., POD) that decomposes $N$ fluctuating velocity fields – $u'_i(\boldsymbol{x},t)$, after removing the ensemble average velocity field from the instantaneous velocity fields – into an optimal basis of $M$ spatial modes and $M$ temporal coefficients ($M \leq N$) according to the following relationship:

$$u'_i(\boldsymbol{x},t) = \sum_m a_m(t)\ \varphi_m(\boldsymbol{x}) \qquad (1)$$

with $a_m$ being an orthogonal basis of temporal coefficients (modal coefficients) and $\varphi_m$ being an orthonormal basis of spatial functions (spatial modes). Here, $t$ and $\boldsymbol{x}$ represent the time and space coordinates vector, respectively.

Coherent structures (the most energetic) are represented by the first modes of the POD. Each mode detected by POD can be described in terms of frequencies, amplitudes, and energy (Sirovich 1987, Berkooz et al., 1993; Pope, 2000). $a_m$ functions incorporate characteristic frequencies ($f_m$) and the energy level captured by each mode ($E_m$) can be obtained by squaring Eq. (1). Given that the velocity fields for an ensemble of $N$ measurements are known, coefficients of Eq. (1) can be obtained following a least-squares procedure.

In the following, $a_m$ are presented for the $m$ first most energetic modes in the ensemble of $N$ velocity fields analysed for each simulation and $f_m$ are the frequencies at which they happen. The code used here is based on the snapshots method (Sirovich, 1987), as implemented by Peltier et al. (2014b).

## 3. Results

### 3.1. Mesh Sensitivity Analysis

Use of the POD allows extraction of the most energetic modes, with their corresponding frequencies and amplitudes. As the analysis is focused on the frequencies, amplitudes and associated energies, mesh sensitivity analysis was also conducted using the POD results instead of other mean flow variables. For completeness of the analysis, results of Peltier et al. (2015) are also included in the following comparisons.

Nine different vertical cell resolutions were explored for mesh sensitivity purposes (Table 2). Mesh sensitivity has been assessed for the two frictional regimes (F and NF), using the RNG *k*-*ε* turbulence model. Given the larger number of simulations performed for case F, discussion will focus on it, whereas similar observations hold for case NF. It is herein assumed that an intermediate level of uncertainty may hold for the transitional cases and that the other considered eddy viscosity models behave likewise.

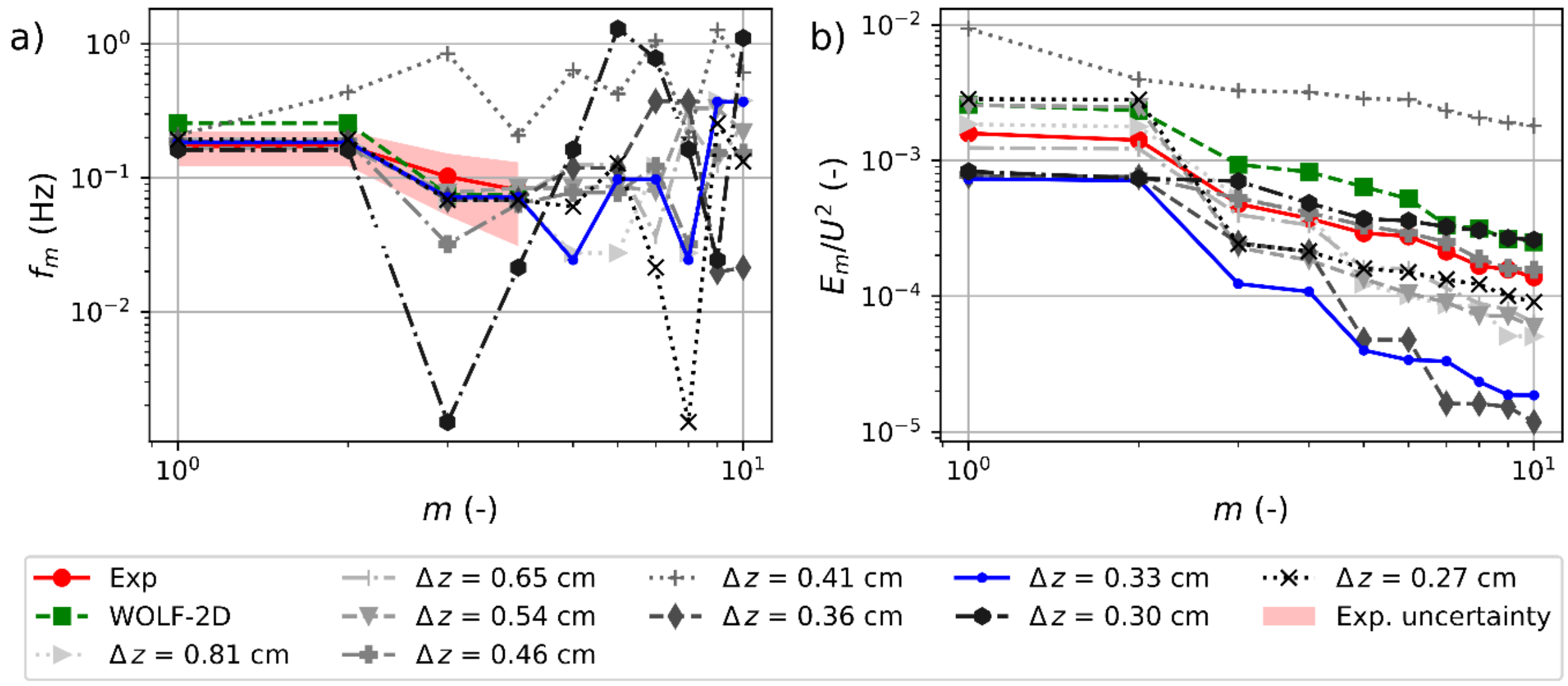

**Figure 2**. Mesh sensitivity analysis for case F: a) modal frequency and b) modal energy; extracted from POD analysis of the ten most energetic modes. Experimental uncertainty up to 0.05 Hz for the frequencies (Peltier et al., 2014b).

Figure 2a shows that most of the tested cell resolutions predicted well the four most energetic frequencies ($f_m$, $m$ = 1, 2, 3 and 4). It can also be observed that minor differences between simulations in the lower modes ($m$) are accompanied of larger differences in higher modes, presenting a large scatter roughly after $m$ = 6. Figure 2b shows the energy predictions for the ten most energetic modes ($E_m$, $m$ = 1 to 10), which hold the biggest part of the flow energy (up to 80 % of the turbulence kinetic energy, according to Peltier et al., 2014b). Figure 2 shows that there is some scatter for different refinement levels, with approximately half the solutions falling above the experimental data and half falling below, intermittently, without a clear dependence on the mesh refinement. Solutions oscillate around a mean level that seems to fall within the experimental uncertainty in the case of frequencies, for which the uncertainty level is known. It is remarkable that both, the coarser and finer meshes, are similarly close to the experimental data, under- and overpredicting the experimental modal energies (Fig. 2b). It is here hypothesized that this variance could be related to small uncontrolled numerical imperfections (for instance, due to the FAVOR obstacle recognition) that lead to early development or detection of first modes at different energy levels. As shown in Fig. 2b, energy is transferred from these modes to the subsequent ones with a reasonably stable rate ($\partial E_m/\partial m$). This energy transfer is similar to the experimental case study, despite the increased eddy viscosity of the simulations. This is consistent with the commonly accepted fact that at low frequencies viscosity plays a negligible effect, thus largest flow structures should not be significantly distorted by the eddy viscosity.

Although the first frequency remains stable with cell refinement, amplitudes ($a_m$) present some oscillations, leading to larger (quadratic) differences in energies (Fig. 2b). Developing vortexes may affect the frequencies of surrounding vortexes and differences are cumulated with increasing modes.

It must be noted that the POD algorithm orders the modes starting from the most energetic one. This does not imply that a similar level of energy at a given mode in a different simulation corresponds spatially to the same vortex. A small increase in the amplitude of a vortex can push it forward in the modal order. The described process may affect the numerical uncertainty that simulations hold (so that there is a mean and variance of the process, being each simulation just a sample). Therefore, classic numerical uncertainty analyses, as for instance that proposed by Celik et al. (2008), would fail to assess the discretization uncertainty. Similar observation holds for case NF, although a more limited number of simulations were run.

The particular mesh resolution herein chosen ($\Delta z$ = 0.33 cm) for the results was not based on the best performance of the simulations – in terms of energies, it undeniably belongs to the group of the lower performing cases – but on the observation that the behaviour remained similar for finer resolutions as well. This cell size was used for all other cases presented in the following results analysis. The vertical discretization for the case F is considerably coarse, yet it has been shown above that no differences were observed with finer meshes. The rest of the cases have larger flow depths, and for a fixed cell size, the number of finite volumes per depth is consequently larger. Note that the mesh sensitivity study presented here involves nine levels of refinement, and it focuses on frequencies and energies of the most energetic modes of the flow. This is beyond common practices, since no previous mesh sensitivity analysis in other Computational Fluid Dynamics studies considers the frequencies and energies of POD modes as indicators of convergence.

### *3.2. Frequencies, Amplitudes and Energy*

Three different turbulence models were tested against the experimental data of Peltier et al. (2014b). For the sake of comparison and analysis, the 2D depth-averaged numerical results of Peltier et al. (2015) and three-dimensional modelling without turbulence viscosity ($\nu_t$ = 0) are also reported herein. All three turbulence models tested seem to perform likewise, with differences probably falling within the uncertainty related to small uncontrolled numerical artefacts.

Figure 3 shows that frequency is generally well reproduced for the four most energetic modes, which is accompanied by a satisfactory result for amplitudes as well (Fig. 4). Larger differences can be observed for the energies (Fig. 5), which scale with the square of the amplitudes.

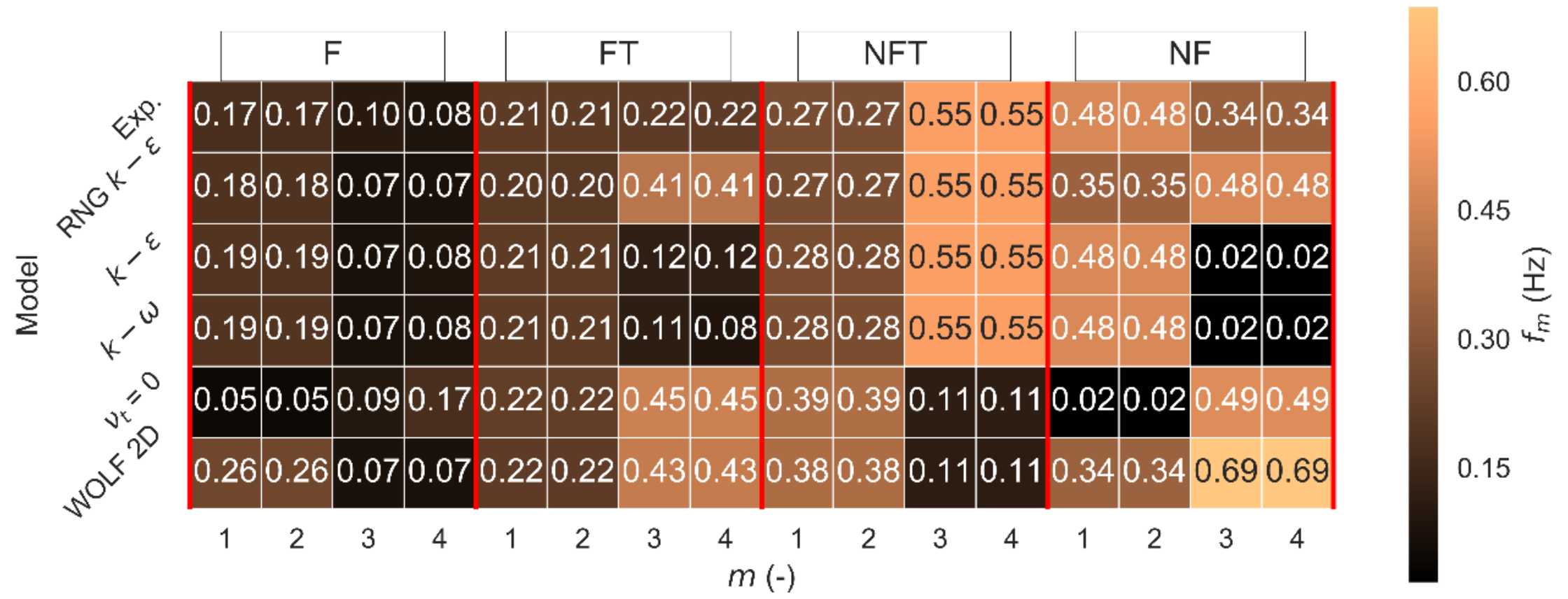

**Figure 3**. Frequency ($f_m$) associated to the four most energetic modes, obtained for cases F, FT, NFT, NF with Δ $z$ = 0.33 cm.

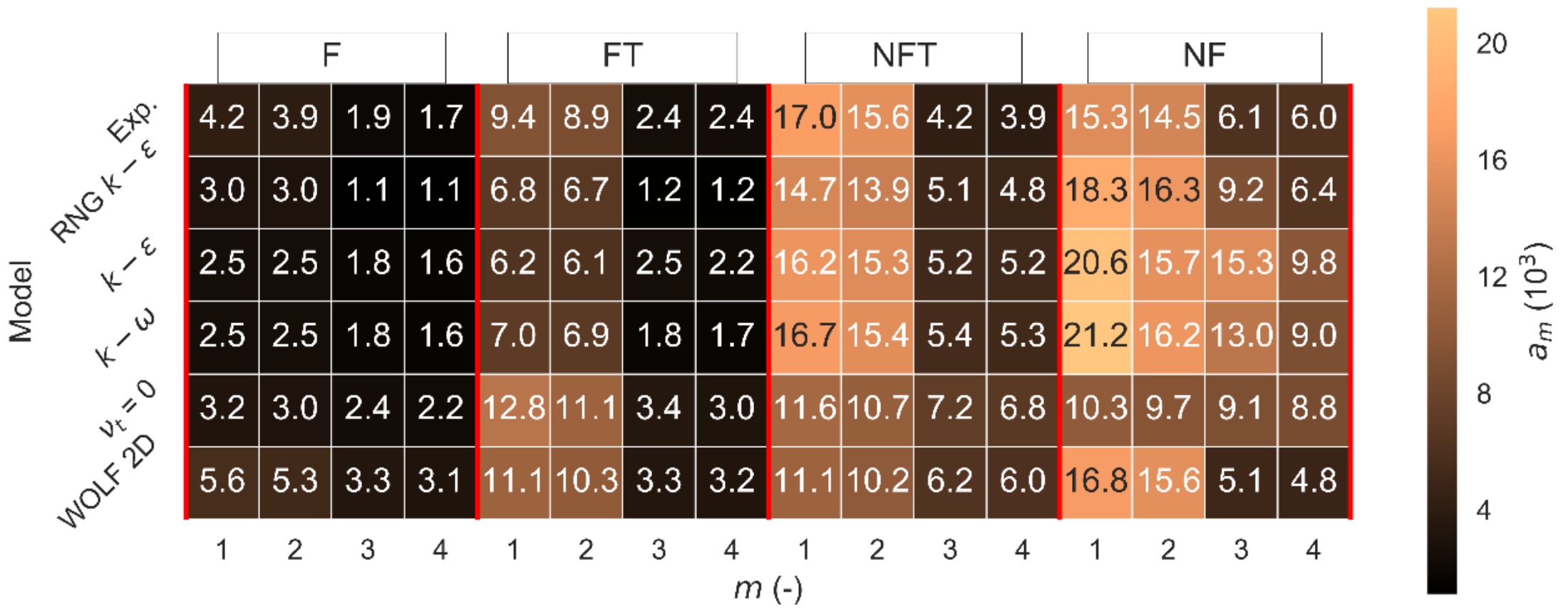

**Figure 4**. Amplitude ($a_m$) associated to the four most energetic modes, obtained for cases F, FT, NFT, NF with Δ $z$ = 0.33 cm.

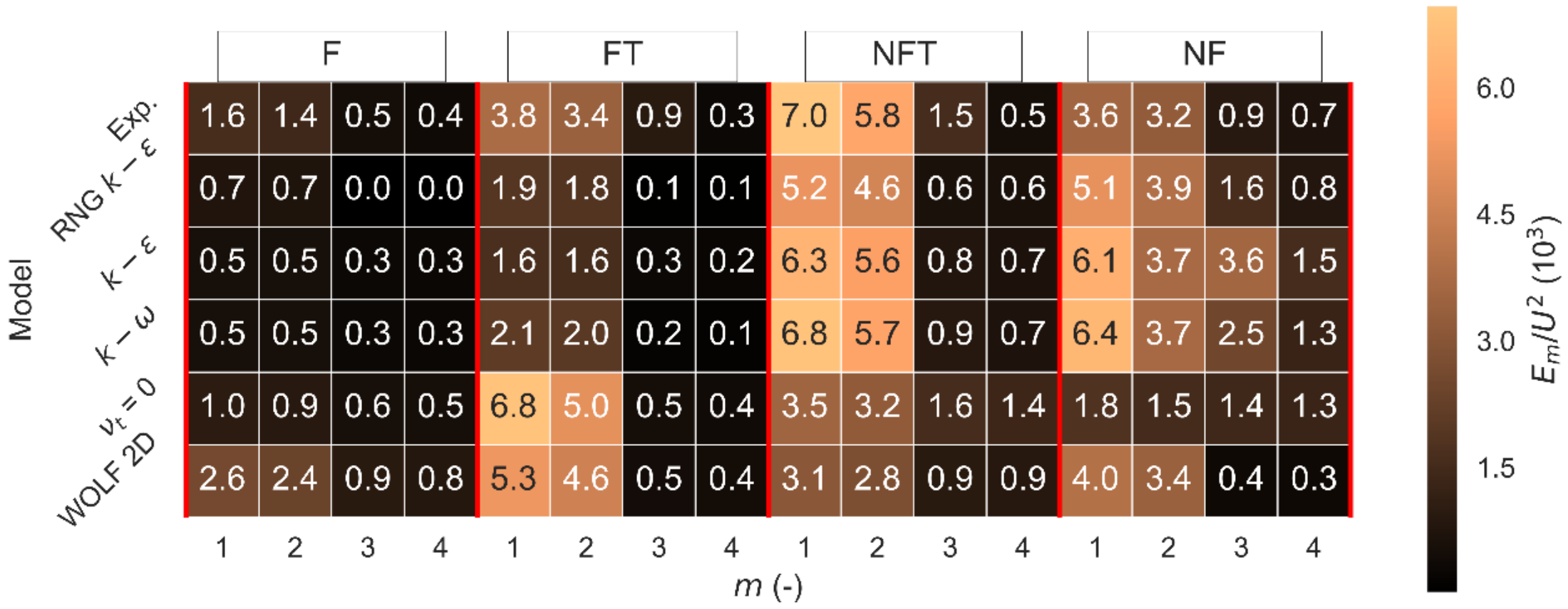

**Figure 5**. Energy ($E_m$) associated to the four most energetic modes, obtained for cases F, FT, NFT, NF with Δ $z$ = 0.33 cm.

Frequencies prediction with the 3D model for cases F and NFT are excellent, with a very good prediction as well of the two most energetic frequencies of cases FT and NF. For $m$ = 3 and 4 of cases FT and NF, the lower performing frequencies, major differences between turbulence models arise. Frequencies associated to $m$ = 1 and 2 of case NF, seem to be exchanged with modes $m$ = 3 and 4 for the RNG $k$-$\varepsilon$ model. Similarly, WOLF 2D and no turbulent viscosity 3D model are able to capture two of the four first modes frequencies, except for case NFT. WOLF 2D results are generally lesser satisfactory than 3D model results, except for $m$ = 1 and 2 in case FT and for $m$ = 3 and 4 in case F. For $m$ = 3 and 4 in case FT and $m$ = 1 and 2 in case NF, WOLF 2D gives similar results to the RNG $k$-$\varepsilon$ model. In other cases, WOLF 2D behaves more like no turbulent viscosity 3D model, except for $m$ = 1 and 2 in case F and $m$ = 3 and 4 in case NF. Since the POD code orders by energies, frequencies could still be in good agreement but allocated to different modes due to differences in energies, thus resulting in a mismatch. It is noteworthy that for the case with no turbulent viscosity, only $m$ = 1 and 2 of case FT and $m$ = 3 in case F are satisfactorily predicted. Hence, eddy viscosity not only helps predicting mean variables, as it is well-known, but it concurrently improves the estimation of fluctuating quantities and does not produce a damping of the frequencies. This is remarkable, as eddy viscosity models are regarded as “steady” models, which have been empirically calibrated against mean variables. Furthermore, the eddy viscosity hypothesis implies that Reynolds stresses principal axis are coinciding with those of the mean strain rate (Wilcox, 2006). This does not hold true for flows with sudden changes in the mean strain rate, strong streamline curvature or secondary currents (Wilcox, 2006). Predictive deterioration of isotropic eddy viscosity models may appear in regions of high turbulence anisotropy (Kang and Sotiropoulos, 2012).

Amplitudes are shown in Fig. 4. It can be observed that with increasing Reynolds number, amplitudes also grow and amplitudes of the first two modes are generally 100 to 400 % larger than third and fourth modes. This is typical for meandering flows due to the interaction of two kinds of instabilities: sinuous and varicose modes (see Peltier et al., 2014b), being the first one responsible of the meandering nature of the flow and the second one of the lateral spreading. The relation between the first two modes and the subsequent ones typically holds for the three tested turbulence models, with some differences arising in case NF for $k$-$\varepsilon$ and $k$-$\omega$ models. It cannot be confirmed that turbulent viscosity produces a damping of the fluctuation intensity as the no turbulent viscosity case occasionally over- or underpredicts the amplitudes further than the three tested turbulence models.

Except for case NF, the amplitudes prediction is commonly improved in the three-dimensional modelling option, despite the flow being markedly two-dimensional. This can be due to 3D velocity profiles developing in confined regions, such as in the developing shear layer immediately in the entry of the basin. This localized phenomenon may impact the downstream development of the large-scale structures and incorporation to 2D models may result complicated. Another explanation could be linked to the three-dimensionality of the large-scale structures. It has been traditionally suggested that in shallow reservoirs the turbulent structures are bounded in the vertical direction by the limited depth. Besides, turbulent structures are subject to bed friction effects. In a shallow flow, the friction is described by the frictional number S and higher values denote increasing relevance of flow resistance on the overall flow (Table 1). With decreasing S, three-dimensionality may play a more significant role and thus turbulent structures should only be considered pseudo-two-dimensional. Results for modal energies can be understood in a similar manner (Fig. 5), regardless of larger differences when compared to the experimental data. Best agreement for the largest Reynolds number case (NF) is obtained using WOLF 2D or RNG $k$-$\varepsilon$ model.

## 4. Discussion

### *4.1. Suitability of Reynolds Averaged Models for Unsteady Flows*

When applying Reynolds Averaging over the flow equations, a certain time smoothing over a large enough time window is applied. Conversely, LES methods use a spatial filtering approach, which allows time to remain free of strict mathematical operations. Rodi (2017) argued that when a URANS model produces low eddy viscosity (given the similarity with LES equations), the URANS calculations could be able to resolve the unsteady motion. However, Rodi (2017) also suggested that frequently eddy viscosity in URANS models is large and damping of turbulent motions is produced. Conversely, the POD analysis in this study showed that no shift is produced in terms of frequencies, and amplitudes are reasonably reproduced for the two most energetic modes. Furthermore, damping and enhancement may happen indifferently (Figs. 4 and 5), which might be consistent with the dual role of viscosity: stabilizing due to dissipation and a subtler destabilizing role (pp 160, Drazin, 2002).

One fundamental limitation of the employed URANS models is that eddy viscosity is commonly considered as a scalar, instead of a tensor. Other available models addressing this issue are Reynolds stress transport models. The gain in universality comes on the expense of adding complexity and closure terms to the model equations (Wilcox, 2006). This handicaps development and calibration of those models and, as Slotnick et al., (2014) argue, they lack robustness and are often less accurate than standard eddy viscosity models.

### *4.2. Performance Assessment*

A relative error ($e$) for a given variable ($\xi$) can be estimated as:

$$e = \frac{\xi - \xi_{\mathrm{Exp}}}{\xi_{\mathrm{Exp}}} \tag{2}$$

with $\xi_{\mathrm{Exp}}$ being the reciprocal, experimentally determined value of variable $\xi$. Correspondingly, a performance/accuracy indicator $\phi(\xi)$ can be expressed as:

$$\begin{aligned} \phi(\xi) &= 1 - |e(\xi)|; \quad \text{for } e(\xi) < 1 \\ \phi(\xi) &= 0; \quad \text{otherwise} \end{aligned} \tag{3}$$

which ranges from 0 to 1 (lowest and highest performance). The reader may note that with Eq. (3) any relative error larger than 100 % is considered to result in $\phi(\xi) = 0$. This accuracy indicator has been computed for all simulated cases and previous WOLF 2D results of Peltier et al. (2015), as shown in Fig. 6. It can be observed that the performance of the 3D model is similar or superior to the two-dimensional model and accuracy remains more stable along all considered cases, except NF. Performance of $k$-$\omega$ is similar to $k$-$\varepsilon$ and both curves regularly overlap. For the RNG $k$-$\varepsilon$ model, results were close to the other two turbulence models with slightly improved performance for the highest Reynolds number case (NF). Yet, differences between the three turbulence models are smaller than differences induced by small meshing changes (Fig. 2) and thus, performance of the three turbulence models should be deemed identical. A combined "mean" (average) result of the three turbulence models is also incorporated in the performance assessment analysis of Fig. 6, highlighting the expected behaviour of a two equations eddy viscosity model (indistinctly of its name). It must be noted that the results presented in this study do not allow generalisation to other cases, but the predictions of the used URANS models consistently showed good accuracy for frequencies, regardless of the turbulence model used.

### *4.3. Three-dimensional flow structure*

Clear differences between the performance of three-dimensional (FLOW-3D©) and two-dimensional modelling (WOLF 2D) were observed, as shown in Fig. 6. Understanding the mechanisms of occurrence of these unstable/turbulent flow cases may help formulating more accurate two-dimensional turbulence models, which in turn can run faster and over larger domains than 3D modelling, allowing the modeller to make an informed decision on the model to be used. Differences may arise due to the vertical velocity component, despite turbulent structures commonly being bounded by the depth. Nonetheless, even coarser resolutions in $z$ – with lower capacity to capture the associated velocities – already showed good capabilities to predict frequencies and energies distributions (Fig. 2a, b) better than the two-dimensional approach. Vertical profiles of mean velocities extracted at different reservoir points, however, showed a typical power law profile and thus, it is of higher interest to check unsteady velocity profiles and, more conveniently, the three-dimensional strain and vorticity fields.

A simple technique allowing detection of vortices is the Q-criterion technique (Hunt et al. 1988). This technique uses the vorticity ($\mathbf{\Omega}$) and strain rate ($\mathbf{S}$) to define the parameter $Q$ (Hunt et al., 1988; Haller, 2005):

$$Q = \frac{1}{2}[|\mathbf{\Omega}|^2 - |\mathbf{S}|^2] \qquad (4)$$

When $Q > 0$, the norm of the vorticity tensor dominates over that of the rate of strain and vortices can be discerned. $Q$ parameter was obtained for the simulated cases and exemplary vortical structures are shown in Fig. 7 and in the supplementary material, both for the highest and lowest Reynolds number cases. It must be noted that results in the previous section were presented for $\Delta z$ = 0.33 cm, whereas $\Delta z$ = 0.27 cm is used together with the Q-criterion technique to render vortexes. Resolution is here changed for the sake of visualization, given that smoother vortex surfaces can be observed. Differences on energy and frequencies between these two simulations were presented in Fig. 2.

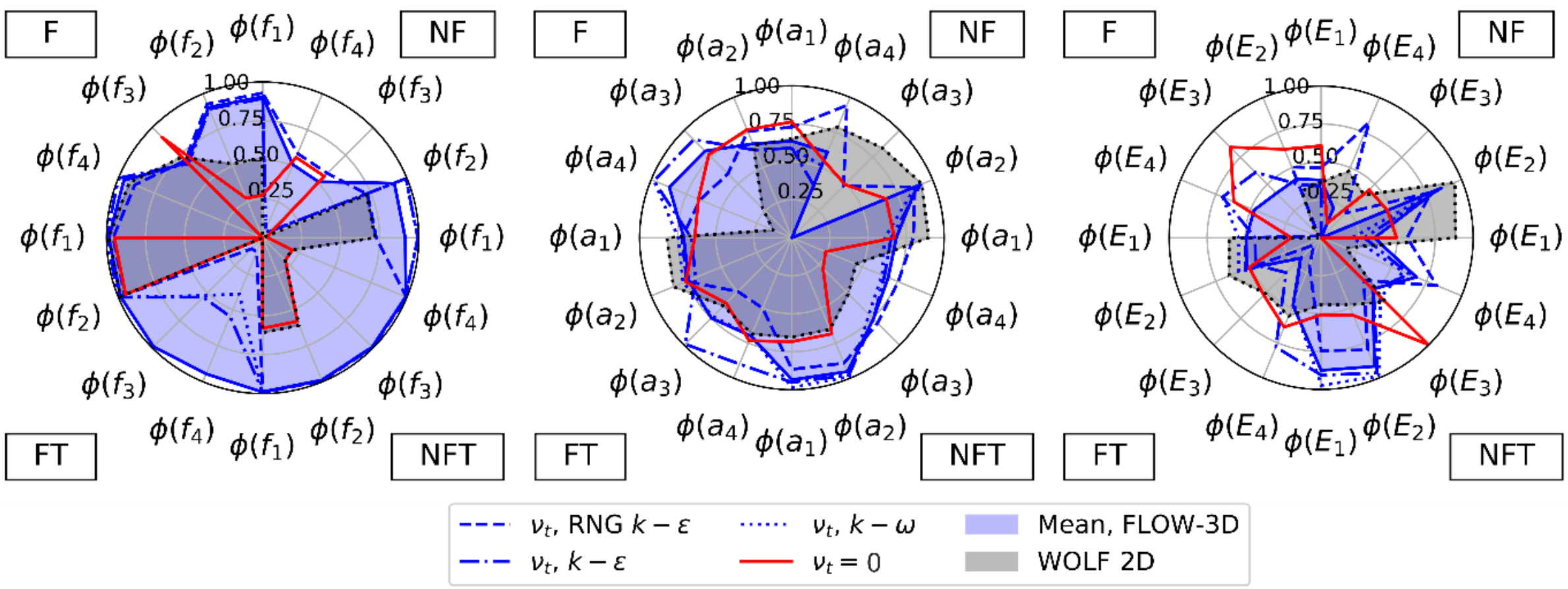

**Figure 6**. Performance assessment (Eq. 3) of the studied three-dimensional turbulence hypotheses and two-dimensional approach (WOLF 2D) of Peltier et al. (2015). "Mean" corresponds to the mean of the results of all three considered turbulence models.

In the inlet channel, a streamwise vortex can be readily observed. The jet entering the basin creates two side regions of large vorticity that insufflate the jet core with small, yet strong, vortical structures that meander around the jet axis. Additionally, transverse velocity profiles had different inflection points at different depths, being this inflection points responsible of instabilities development. Thus, three-dimensional process is a more complex process than the depth-averaged one. Some vortices crawl against the free surface (where the POD was applied) whereas others vanish, never reaching it. Vortices have a three-dimensional body and the centre of gravity could be located at different depths, arbitrarily far from the free surface. The resulting two-dimensional vortical structures occurring in the free surface plane are, simply, the footprint of a more complex phenomenon occurring under the surface; which cannot be directly captured by a depth-averaged model. This could explain limitations of two-dimensional modelling in capturing some pseudo-two-dimensional turbulent structures.

Naturally, stronger fluctuations occur in the case NF than F as, in the former case, shallowness prevents lateral spreading of the turbulent structures. Intensity of the vortices decreases through the reservoir. For the higher Reynolds number case (NF), it can also be observed that vortexes accumulate upstream of the outlet, which is not able to flush them out and some are fed to the recirculation regions while others are stretched around the basin outlet edges, recurrently increasing their intensity. The vortex stretching in the outlet is produced intermittently, at left and right sides, following the inner dynamics of the basin.

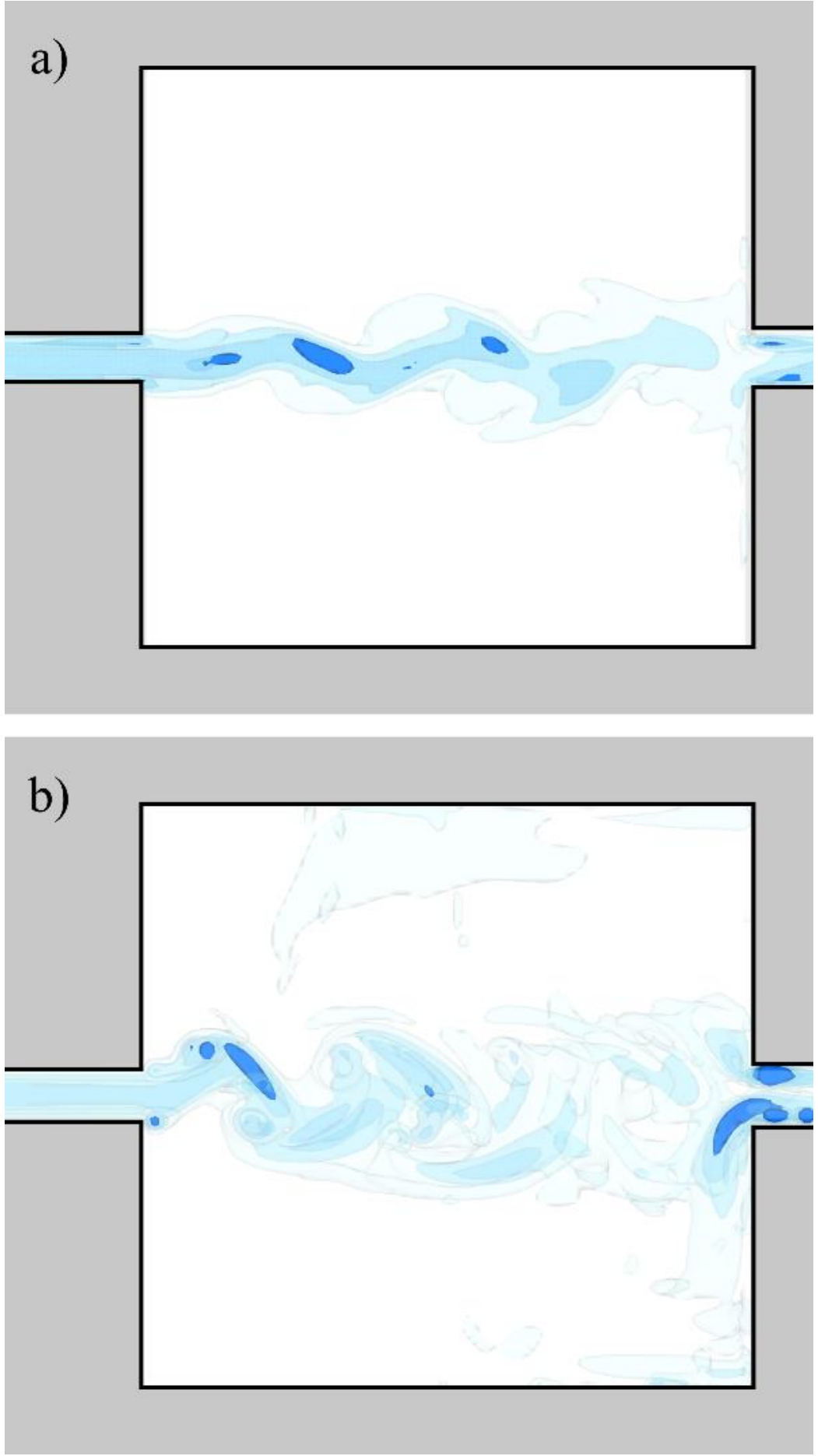

**Figure 7**. Large-eddy identification with the Q-criterion for RNG *k*-ε, Δ *z* = 0.27 cm and cases a) F and b) NF. The different colours correspond to different *Q* isosurfaces, from darker to clearer a) $Q$ = 40, 20, 12.5, 2.5 Hz$^2$ and b) $Q$ = 200, 100, 50, 10 Hz$^2$.

## 5. Conclusions

Four different meandering flow conditions (F, FT, NFT, NF of Table 1) were studied numerically in a shallow rectangular reservoir with the purpose of assessing the performance of three-dimensional URANS together with three turbulence models (RNG $k$-$\varepsilon$, $k$-$\varepsilon$ and $k$-$\omega$). A POD technique was used to extract modal frequencies, amplitudes and energies associated to the most energetic modes. They were compared to a previous experimental study of Peltier et al. (2014b) and 2D depth-averaged numerical results presented by Peltier et al. (2015). The frequencies computed with the three-dimensional models, ensemble-averaged, for the four most energetic modes reproduced 88.7 % of experimental data, while some differences arise for the predicted amplitudes and even to a greater extent for the energies, reproducing respectively 70.7 % and 47.7 % of the experimental results. The three considered turbulence models showed similar performance, with improvements in the predictions when compared to 2D modelling (Fig. 6), that remained at 48.4 %, 65.1 %, 46.3 % for the frequencies, amplitudes and energies. It is noteworthy that the mesh resolution chosen ($\Delta z$ = 0.33 cm) was not based on the best performance of the simulations; undoubtedly, it is representative of the group of the lower performing cases (Fig. 2a,b).

Use of three-dimensional modelling allowed capturing localized effects that remained undetected in previous two-dimensional studies, and that can impact downstream development of 2D large-scale structures. Q-criterion was used to identify vortexes and describe their dynamics (Fig. 7), elucidating on the complexity of the flow. It was found that the three-dimensional flow produces a more intermittent two-dimensional footprint in the free surface, where the POD was conducted. Also, other mechanisms such as 3D vortex stretching may affect the overall unsteady motion of the shallow reservoir.

Here, the focus was set on an engineering perspective, by comparing a widely used 3D modelling strategy (URANS, mesh design, turbulence closures) to previous 2D results. Future research should investigate the ability of more advanced 3D modelling approaches in resolving the flow variables.

**Conflict of interest**

Declarations of interest: none

**Funding**

This research did not receive any specific grant from funding agencies in the public, commercial, or not-for-profit sectors.

**Notation**

*The following symbols are used in this paper:*

*Acronyms:*

CFD Computational Fluid Dynamics

F Frictional regime

FT Frictional close to transition regime

LES Large Eddy Simulation

NF Non-frictional regime

NFT Non-frictional close to transition regime

POD Proper Orthogonal Decomposition

SF Shape Factor

URANS Unsteady Reynolds-Averaged Navier-Stokes

VOF Volume of Fluid

*Roman:*

$a_m$ Amplitude associated to mode $m$

$B$ Width of the shallow reservoir

$b$ Width of the inlet channel

$E_m$ Energy associated to mode $m$

$e$ Relative error

F Froude number

$f_m$ Frequency associated to mode $m$

$H$ Flow depth

$k$ Turbulence kinetic energy

$L$ Length of the shallow reservoir

$M$ Number of spatial modes and temporal coefficients

$m$ Mode obtained with the POD technique

$N$ Number of fluctuating velocity fields

$\mathcal{Q}$ Q-criterion parameter

$Q$ Flow rate

R Reynolds number

$\mathbf{S}$ Strain rate tensor

S Frictional number

$T$ Time window

$t$ Time

$U$ Depth-averaged velocity

$u'_i$ Instantaneous velocity fluctuation

$\boldsymbol{x}$ Coordinate vector

*Greek:*

$\Delta B$ Width of the sudden expansion

$\Delta z$ Cell size resolution in the depth direction

$\varepsilon$ Turbulence dissipation

$\nu_t$ Eddy (turbulent) viscosity

$\xi$ Numerically obtained variable

$\xi_{\text{Exp}}$ Experimental value of the numerical variable $\xi$

$\phi$ Performance indicator

$\varphi_m$ Empirical basis

$\mathbf{\Omega}$ Vorticity tensor

$\omega$ Specific dissipation

**Supplemental material**

Large-eddy identification using the Q-criterion for RNG $k$-$\varepsilon$, $\Delta z$ = 0.27 cm.

Video S1. Case F. $Q$ isosurface colours correspond to: $Q$ = 40, 20, 12.5, 2.5 Hz$^2$.

Video S2. Case NF. $Q$ isosurface colours correspond to: $Q$ = 200, 100, 50, 10 Hz$^2$.

**Data Availability Statement**

Some or all data, models, or code that support the findings of this study are available from the corresponding author upon reasonable request.

**References**

Abbasi, A., Annor, F.O. and Van de Giesen, N., 2016. Investigation of temperature dynamics in small and shallow reservoirs, case study: Lake Binaba, Upper East Region of Ghana. *Water*, 8(3), 84. DOI: 10.3390/w8030084

Adamsson, Å., Stovin, V., Bergdahl, L., 2003. Bed shear stress boundary condition for storage tank sedimentation. Journal of Environmental Engineering, 129(7), 651-658. DOI: 10.1061/(ASCE)0733-9372(2003)129:7(651)

Aloui, F., Souhar, M., 2000. Experimental study of turbulent asymmetric flow in a flat duct symmetric sudden expansion. J. Fluid. Eng., 122(1), 174–177.

Berkooz, G., Holmes, P., Lumley, J.L., 1993. The proper orthogonal decomposition in the analysis of turbulent flows. Annual Review of Fluid Mechanics, 25(1), 539-575. DOI: 10.1146/annurev.fl.25.010193.002543

Buffin-Bélanger, T., Roy, A.G., Demers, S., 2013. Turbulence in River Flows. In: Treatise in Geomorphology, 1st edition, Editor: Ellen Wohl. Elsevier. DOI: 10.1016/B978-0-12-374739-6.00231-1

Camnasio, E., Orsi, E., Schleiss, A.J., 2011. Experimental study of velocity fields in rectangular shallow reservoirs. J. Hydraul. Res., 49(3), 352-358. DOI: 10.1080/00221686.2011.574387

Camnasio, E., Erpicum, S., Orsi, E., Pirotton, M., Schleiss, A.J., Dewals, B., 2013. Coupling between flow and sediment deposition in rectangular shallow reservoirs. J. Hydraul. Res., 51, 535–547.

Camnasio, E., Erpicum, S., Archambeau, P., Pirotton, M., Dewals, B. 2014. Prediction of mean and turbulent kinetic energy in rectangular shallow reservoirs. Eng. Appl. Comput. Fluid Mech., 8,586–597.

Canbazoglu, S., Bozkir, O., 2004. Analysis of pressure distribution of turbulent asymmetric flow in a flat duct symmetric sudden expansion with small aspect ratio. Fluid Dyn. Res., 35(5), 341–355.

Celik, I.B., Ghia, U., Roache, P.J., 2008. Procedure for estimation and reporting of uncertainty due to discretization in CFD applications. Journal of Fluids Engineering, 130 (7): 1–4. DOI: 10.1115/1.2960953

Chen, D., Jirka, G.H., 1999. LIF study of plane jet bounded in shallow water layer. J. Hydraul. Eng., 10.1061/(ASCE)0733-9429 (1999)125:8(817), 817–826.

Dewals, B.J., Kantoush, S.A., Erpicum, S., Pirotton, M., Schleiss, A.J., 2008. Experimental and numerical analysis of flow instabilities in rectangular shallow basins. Environmental Fluid Mechanics, 8(1), 31-54. DOI: 10.1007/s10652-008-9053-z

Drazin, P.G., 2002. Introduction to hydrodynamic stability. Cambridge University Press.

Dufresne, M., Dewals, B.J., Erpicum, S., Archambeau, P., Pirotton, M., 2010a. Classification of flow patterns in rectangular shallow reservoirs. Journal of Hydraulic Research, 48(2), 197-204. DOI: 10.1080/00221681003704236

Dufresne, M., Dewals, B., Erpicum, S., Archambeau, P., Pirotton, M., 2010b. Experimental investigation of flow pattern and sediment deposition in rectangular shallow reservoirs. International Journal of Sediment Research, 25(3), 258-270. DOI: 10.1016/S1001-6279(10)60043-1

Dufresne, M., Dewals, B.J., Erpicum, S., Archambeau, P., Pirotton, M., 2011. Numerical investigation of flow patterns in rectangular shallow reservoirs. Eng. Appl. Comput. Fluid Mech., 5(2), 247–258.

Erpicum, S., Meile, T., Dewals, B.J., Pirotton, M., Schleiss, A.J., 2009. 2D numerical flow modeling in a macro-rough channel. International Journal for Numerical Methods in Fluids, 61(11), 1227-1246. DOI: 10.1002/fld.2002

Esmaeili, T., Sumi, T., Kantoush, S.A., Haun, S., Rüther, N., 2016. Three-dimensional numerical modelling of flow field in shallow reservoirs. Proceedings of the Institution of Civil Engineers-Water Management, 169(5), 229-244. DOI: 10.1680/jwama.15.00011

Franca, M. J., Valero, D., Liu, X., 2021. Turbulence and rivers. In: Treatise in Geomorphology, 2nd edition, Editor: Ellen Wohl. Elsevier. DOI: 10.1016/B978-0-12-818234-5.00135-8

Ge, L., Sotiropoulos, F., 2005. 3D unsteady RANS modeling of complex hydraulic engineering flows. I: Numerical model. Journal of Hydraulic Engineering, 131(9), 800-808. DOI: 10.1061/(ASCE)0733-9429(2005)131:9(800)

Haller, G., (2005). An objective definition of a vortex. Journal of Fluid Mechanics, 525, 1-26. DOI: 10.1017/S0022112004002526

Haun, S., Kjærås, H., Løvfall, S. and Olsen, N.R.B., 2013. Three-dimensional measurements and numerical modelling of suspended sediments in a hydropower reservoir. *Journal of Hydrology*, 479, 180-188. DOI: 10.1016/j.jhydrol.2012.11.060

Haun, S. and Olsen, N.R.B., 2012. Three-dimensional numerical modelling of reservoir flushing in a prototype scale. *International Journal of River Basin Management*, 10(4), 341-349. DOI: 10.1080/15715124.2012.736388

Hirt, C.W., Nichols, B.D., 1981. Volume of fluid (VOF) method for the dynamics of free boundaries. Journal of Computational Physics, 39(1), 201-225. DOI: 10.1016/0021-9991(81)90145-5

Hirt, C.W., Sicilian, J.M., 1985. A porosity technique for the definition of obstacles in rectangular cell meshes. Proc.: 4th International Conference on Numerical Ship Hydrodynamics.

Hunt, J.C., Wray, A.A., Moin, P., 1988. Eddies, streams, and convergence zones in turbulent flows. NASA Technical Report.

Isenmann, G., Dufresne, M., Vazquez, J., Mose, R., 2017. Bed turbulent kinetic energy boundary conditions for trapping efficiency and spatial distribution of sediments in basins." Water Science and Technology, wst2017373. DOI: 10.2166/wst.2017.373

Kang, S., Sotiropoulos, F., 2012. Assessing the predictive capabilities of isotropic, eddy viscosity Reynolds-averaged turbulence models in a natural-like meandering channel. Water Resour. Res., 48, W06505, DOI:10.1029/2011WR011375

Kantoush, S.A., De Cesare, G., Boillat, J.L., Schleiss, A.J., 2008. Flow field investigation in a rectangular shallow reservoir using UVP, LSPIV and numerical modelling. Flow measurement and Instrumentation, 19(3-4), 139-144. DOI: 10.1016/j.flowmeasinst.2007.09.005

Khan, S., Melville, B.W., Shamseldin, A.Y., Fischer, C., 2013. Investigation of flow patterns in storm water retention ponds using CFD. J. Environ. Eng., DOI: 10.1061/(ASCE)EE.1943-7870.0000540, 61–69.

Khosronejad, A., Kang, S., Sotiropoulos, F., 2012. Experimental and computational investigation of local scour around bridge piers. Advances in Water Resources, 37, 73-85.

Liu, X., Xue, H., Hua, Z., Yao, Q., Hu, J., 2013. Inverse calculation model for optimal design of rectangular sedimentation tanks. Journal of Environmental Engineering, 139, 455–459.

Mullin, T., Shipton, S., Tavener, S.J., 2003. Flow in a symmetric channel with an expanded section. Fluid Dyn. Res., 33(5–6), 433–452.

Oca, J., Masaló, I., (2007). Design criteria for rotating flow cells in rectangular aquaculture tanks. Aquacultural Engineering, 36, 36–44.

Palkin, E., Mullyadzhanov, R., Hadžiabdić, M., Hanjalić, K., 2016. Scrutinizing URANS in shedding flows: the case of cylinder in cross-flow in the subcritical regime. Flow, Turbulence and Combustion, 97(4), 1017-1046. DOI: 10.1007/s10494-016-9772-z

Peltier, Y., Erpicum, S., Archambeau, P., Pirotton, M., Dewals, B., 2014a. Experimental investigation of meandering jets in shallow reservoirs. Environmental Fluid Mechanics, 14(3), 699-710. DOI: 10.1007/s10652-014-9339-2

Peltier, Y., Erpicum, S., Archambeau, P., Pirotton, M., Dewals, B., 2014b. Meandering jets in shallow rectangular reservoirs: POD analysis and identification of coherent structures. Experiments in Fluids, 55(6), 1740. DOI: 10.1007/s00348-014-1740-6

Peltier, Y., Erpicum, S., Archambeau, P., Pirotton, M., Dewals, B., 2015. Can meandering flows in shallow rectangular reservoirs be modeled with the 2D shallow water equations?. Journal of Hydraulic Engineering, 141(6), 04015008. DOI: 10.1061/(ASCE)HY.1943-7900.0001006

Peng, Y., Zhou, J.G., Burrows, R., 2011. Modeling Free-Surface Flow in Rectangular Shallow Basins by Using Lattice Boltzmann Method. Journal of Hydraulic Engineering, 137(12), 1680-1685. DOI: 10.1061/(ASCE)HY.1943-7900.0000470

Pope, S.B., 2000. Turbulent Flows. Cambridge University Press.

Rodi, W., 2017. Turbulence modeling and simulation in hydraulics: a historical review. Journal of Hydraulic Engineering, 143(5), 03117001. DOI: 10.1061/(ASCE)HY.1943-7900.0001288

Saad, Y., 2003. Iterative methods for sparse linear systems. Siam, Second Edition. ISBN: 978-0-898715-34-7

Sébastian, C., Becouze-Lareure, C., Kouyi, G. L., Barraud, S., 2015. Event-based quantification of emerging pollutant removal for an open stormwater retention basin–Loads, efficiency and importance of uncertainties. Water Research, 72, 239-250. DOI: 10.1016/j.watres.2014.11.014

Sirovich, L., 1987. Turbulence and the dynamics of coherent structures. Part I: coherent structures. Quarterly of Applied Mathematics 45, 561-570.

Slotnick, J., Khodadoust, A., Alonso, J., Darmofal, D., Gropp, W., Lurie, E., Mavriplis, D., 2014. CFD vision 2030 study: a path to revolutionary computational aerosciences. NASA/CR–2014-218178

Spalart, P.R., 2000. Strategies for turbulence modelling and simulations. International Journal of Heat and Fluid Flow, 21(3), 252-263. DOI: 10.1016/S0142-727X(00)00007-2

Stovin, V.R., Saul, A.J., 2000. Computational fluid dynamics and the design of sewage storage chambers. Water and Environment Journal, 14(2), 103-110. DOI: 10.1111/j.1747-6593.2000.tb00235.x

Valero, D. and Bung, D.B., 2016. Sensitivity of turbulent Schmidt number and turbulence model to simulations of jets in crossflow. Environmental Modelling & Software, 82, 218-228. DOI: 10.1016/j.envsoft.2016.04.030

Valero, D., Bung, D.B., Crookston, B.M., 2018. Energy dissipation of a Type III basin under design and adverse conditions for stepped and smooth spillways. Journal of Hydraulic Engineering, 144(7), p.04018036. DOI: 10.1061/(ASCE)HY.1943-7900.0001482

Valero, D., Bung, D.B., Erpicum, S., Dewals, B., 2017. Numerical study of turbulent oscillations around a cylinder: RANS capabilities and sensitivity analysis. Proc.: 37th IAHR World Congress, Kuala Lumpur, Malaysia, 13 – 18 August, 2017.

Van Leer, B., 1977. Towards the ultimate conservative difference scheme. IV. A new approach to numerical convection. Journal of Computational Physics, 23(3), 276-299. DOI: 10.1016/0021-9991(77)90095-X

Versteeg, H.K., Malalasekera, W., 2007. An Introduction to Computational Fluid Dynamics: The Finite Volume Method. Pearson Education, Second Edition. ISBN: 978-0131274983

Westhoff, M.C., Erpicum, S., Archambeau, P., Pirotton, M., Dewals, B., 2018. Maximum energy dissipation to explain velocity fields in shallow reservoirs. Journal of Hydraulic Research, 56(2), 221-230. DOI: 10.1080/00221686.2017.1289268

Wilcox, D.C., 2006. Turbulence modeling for CFD. DCW Industries, Third Edition. ISBN: 978-1-928729-08-2

Zhang, J.-M., Lee, H.P., Khoo, B.C., Peng, K.Q., Zhong, L., Kang, C.-W. Ba, T., 2014. Shape effect on mixing and age distributions in service reservoirs. Journal - American Water Works Association, 106, E481–E491.